\documentclass[twocolumn]{aastex62}
\usepackage{times}
\usepackage{amsmath}
\usepackage{graphicx}
\usepackage{hyperref}
\usepackage{gensymb}
\usepackage{upgreek}
\usepackage{natbib}
\usepackage{subfigure}
\usepackage{multirow}
\usepackage{comment}
\usepackage{mathtools}
\usepackage{mathrsfs}
\usepackage{fontenc}
\usepackage{color}
\usepackage{url}
\usepackage{pifont}

\newcommand{\HII}{H\,\textsc{ii}}
\newcommand{\Ha}{H$\alpha$}
\newcommand{\Hb}{H$\beta$}
\newcommand{\NeII}{[Ne\,\textsc{ii}]}
\newcommand{\NeIII}{[Ne\,\textsc{iii}]}
\newcommand{\NeV}{[Ne\,\textsc{v}]}
\newcommand{\OII}{[O\,\textsc{ii}]}
\newcommand{\OIII}{[O\,\textsc{iii}]}

\newcommand{\NII}{[N\,\textsc{ii}]}

\submitjournal{ApJ}

\shorttitle{\OII\ SFR Estimator}

\shortauthors{Zhuang \& Ho}

\begin{document}

\title{Recalibration of \OII\ $\lambda 3727$ as a Star Formation Rate Estimator for Active and Inactive Galaxies}

\author[0000-0001-5105-2837]{Ming-Yang Zhuang}
\email{mingyangzhuang@pku.edu.cn}
\affil{Kavli Institute for Astronomy and Astrophysics, Peking University,
Beijing 100871, China}
\affil{Department of Astronomy, School of Physics, Peking University,
Beijing 100871, China}

\author[0000-0001-6947-5846]{Luis C. Ho}
\affil{Kavli Institute for Astronomy and Astrophysics, Peking University,
Beijing 100871, China}
\affil{Department of Astronomy, School of Physics, Peking University,
Beijing 100871, China}

\begin{abstract}
We investigate the use of the \OII\ $\lambda3727$ emission line as a star 
formation rate (SFR) estimator using Sloan Digital Sky Spectra for nearly 
100,000 star-forming galaxies and 5,500 galaxies with narrow-line active 
galactic nuclei.  Consistent with previous work, we find that the \OII/\Ha\ 
ratio in star-forming galaxies depends strongly on gas-phase metallicity.  
Using metallicities derived from the \NII\ $\lambda 6584$/\OII\ $\lambda 3727$ 
method, we refine a metallicity-dependent SFR estimator based on \OII\ that is 
calibrated within a scatter of 0.056 dex against the more commonly used SFR 
indicator based on \Ha\ emission.  The scatter increases to only 0.12 dex if the 
metallicity is estimated using the stellar mass-metallicity relation.  With the aim 
of extending the \OII-based SFR estimator to active galaxies, we calculate radiation 
pressure-dominated photoionization models to constrain the amount of \OII\ 
emission arising from the narrow-line region.  We use the sample of active 
galaxies to demonstrate that the SFRs derived from \OII, after accounting for 
nonstellar contamination, are consistent with independent SFR diagnostics 
estimated from the stellar continuum of the host galaxies. 
\end{abstract}


\section{Introduction} \label{sec1}

The star formation rate (SFR) is one of the most fundamental physical parameters
to understand the formation and evolution of galaxies. A variety of indicators 
have been calibrated to estimate the SFR in star-forming galaxies, ranging from
ultraviolet, infrared, and radio continuum emission to emission lines tracing
photoionized or photodissociated regions \citep[for a review, see][]{2012ARA&A..50..531K}. 
The nebular recombination line \Ha\ $\lambda6563$ is typically 
regarded as the most reliable SFR indicator.  Proportional to the ionizing 
radiation from young ($\lesssim 20$ Myr), massive ($>10 \ M_{\odot}$) stars, 
the bright and widely accessible \Ha\ line provides a direct probe of the 
instantaneous SFR independent of previous star formation history 
\citep{1998ARA&A..36..189K}.  Beyond $z \approx 0.5$, when \Ha\ is redshifted 
outside of the optical window, the most widely considered alternative 
spectroscopic SFR indicator is the \OII\ $\lambda3727$ doublet 
\citep[e.g.,][]{1989AJ.....97..700G, 1992ApJ...388..310K, 2004AJ....127.2002K, 
2006ApJ...642..775M, 2007ApJ...660L..39W, 2009A&A...495..759A, 
2010MNRAS.405.2594G}, which can extend the redshift coverage up to $z \approx 
1.7$.  However, \OII\ is less directly connected with the ionizing photons 
than \Ha, and several complicating factors need to be considered. 

\citet{2004AJ....127.2002K} used a nearby galaxy sample to study the variation 
of the \OII/\Ha\ ratio with dust reddening, metallicity, and ionization 
parameter, and offered an improved \OII\ SFR calibration with an explicit 
correction for metallicity. \citet{2006ApJ...642..775M} also investigated the 
systematic uncertainties of the \OII-based SFR diagnostic, incorporating 
$B$-band luminosity as a term to reduce the scatter due to dust reddening, 
metallicity, and ionization. Other improvements to the \OII\ SFR indicator have
been proposed in a similar spirit \citep{2007ApJ...660L..39W, 
2009A&A...495..759A, 2009ApJ...703.1672K, 2010MNRAS.405.2594G}.  However, 
these modified calibrations are still subject to some shortcomings, such as 
relatively small calibration sample, large scatter ($\sim$0.3 dex), or the 
requirement of additional observations. 

Ever since the recognition that black hole mass correlates tightly with galaxy 
properties \citep{1998AJ....115.2285M, 2000ApJ...539L...9F, 2000ApJ...539L..13G, 
2013ARA&A..51..511K}, much attention has been devoted to the manner in which 
supermassive black holes might coevolve with their host galaxies 
\citep[e.g.,][]{1998Natur.395A..14R,2004coa..book.....H,2014ARA&A..52..589H}.
A key hindrance to progress in this still-controversial subject stems from the 
fact that reliable SFRs are tremendously difficult to ascertain in active 
galactic nuclei (AGNs).  AGN emission encompasses the entire spectral energy 
distribution (SED), presenting an unavoidable source of contamination for 
virtually all traditional extragalactic SFR estimators.  In terms of 
emission-line diagnostics arising from photoionized gas, a strategy must be 
devised to separate the contribution from the AGN narrow-line region (NLR) to 
that arising from \HII\ regions.  \citet{2019ApJ...873..103Z} recently 
presented a new method to derive SFRs in active galaxies using the mid-IR 
fine-structure lines \NeII\ 12.81 \micron\ and \NeIII\ 15.55 \micron, which 
effectively trace the ionizing luminosity of star-forming galaxies \citep
{2007ApJ...658..314H}.  While the NLR of AGNs also emits copious \NeII\ and 
\NeIII, in the case of nonstellar excitation these low-ionization lines are 
unavoidably accompanied by \NeV\ 14.32 \micron\ for the ionization parameters 
characteristic of Seyfert galaxies and quasars.  Fortunately, the ratio of 
\NeII\ or \NeIII\ relative to \NeV\ spans a sufficiently well-defined, narrow 
range that the contribution of the AGN to the low-ionization lines can be 
removed, and hence the SFR of the underlying host inferred.  Zhuang et al.'s 
methodology for treating the neon lines can also be adopted for \OII\ 
$\lambda 3727$, whose intensity relative to \OIII\ $\lambda 5007$ also 
occupies a fairly restricted range in the NLR of highly accreting AGNs 
\citep{2005ApJ...629..680H, 2006ApJ...642..702K}.
In both cases, the lines of low ionization potential (\NeII, 
\NeIII, and \OII) can be excited by both star formation and nuclear activity, 
while the lines of high ionization potential (\NeV\ and \OIII) are powered 
predominantly by the AGN.

Here we present a recalibration of \OII\ as a SFR indicator using nearly
100,000 star-forming galaxies, the largest sample to date, drawn from the 
seventh data release \citep[DR7;][]{2009ApJS..182..543A} of the Sloan Digital 
Sky Survey \citep[SDSS;][] {2000AJ....120.1579Y}.  The large sample size 
enables us to probe in greater detail than previous work the dependence of the 
\OII/\Ha\ ratio on oxygen abundance, achieving a much better global consistency
between the \OII\ and \Ha\ SFR indicators (0.056 dex scatter). We compute a 
suite of new NLR models, spanning a wide range of realistic physical conditions,
to constrain the amount of \OII\ at a given strength of \OIII\ produced by AGNs.
These models enables us to isolate the contribution of nonstellar 
photoionization to the observed, total integrated \OII\ emission of an active 
galaxy, thereby separating the fraction of the line attributable to 
star-forming regions in the host galaxy.  We apply our new \OII\ SFR 
calibration to $\sim$5,500 narrow-line AGNs and show its consistency with an 
independent SFR diagnostic.

Section \ref{sec2} describes the data and their selection criteria. We calibrate
\OII\ as a SFR estimator for star-forming galaxies in Section \ref{sec3}, and 
then in Section \ref{sec4} we extend it to AGNs with the help of NLR 
photoionization models. Section \ref{sec5} discusses the results from our models
and compares them with independent SFRs for AGNs.  Our main conclusions are 
summarized in Section \ref{sec6}. Throughout the paper, we assume a cosmology 
with $H_0=70$ km s$^{-1}$ Mpc$^{-1}$, $\Omega_m=0.3$, and $\Omega_{\Lambda}
=0.7$.  As in \citet{2004MNRAS.351.1151B}, we adopt the \citet
{2001MNRAS.322..231K} stellar initial mass function.  For the initial 
mass functions of either \citet{1955ApJ...121..161S} or \citet{2003PASP..115..763C}, 
our SFRs need to be scaled by a factor of 1.49 and 0.94, respectively 
\citep{2014ARA&A..52..415M}.

\section{Data} \label{sec2}

Our data are drawn from SDSS DR7, which covers an area of 8423 deg$^2$, with 
spectroscopy of complete samples of galaxies and quasars covering $\sim$8200 
deg$^2$. The spectra are taken with a $3^{\prime\prime}$-diameter fiber with 
spectral coverage 3800$-$9200 \AA\ at a resolution of $\lambda/\Delta\lambda 
\approx 2000$.  We use the emission-line fluxes \citep{2004ApJ...613..898T} 
and stellar mass ($M_*$) measurements \citep{2003MNRAS.341...33K, 
2007ApJS..173..267S} provided by the Max Planck Institute for Astrophysics 
and Johns Hopkins University (MPA-JHU) catalog\footnote
{http://www.strw.leidenuniv.nl/$\sim$jarle/SDSS/, 
http://www.mpa-garching.mpg.de/SDSS/DR7}.  
We adopt the nuclear spectral classifications from the MPA-JHU catalog, which 
are based on the precepts of \citet{1981PASP...93....5B} discussed in \citet
{2004MNRAS.351.1151B}.  We correct the line fluxes for dust extinction using 
the observed Balmer decrements and the Milky Way extinction curve of \citet
{1989ApJ...345..245C}\footnote{Most of the objects in our final sample have 
mild reddening, with a median $A_V$ of 0.88 mag.} with $R_V=A_V/E(B-V)=3.1$.
For electron temperatures 
$T_e=10^4$ K and electron densities $n_e\approx10^2-10^4$ cm$^{-3}$, the 
intrinsic value of \Ha/\Hb\ $= 2.86$ for star-forming galaxies and \Ha/\Hb\ $= 3.1$ 
for AGNs \citep{2006agna.book.....O}. 

The MPA-JHU catalog lists 203,630 star-forming galaxies and 91,477 
AGNs\footnote{We exclude low-ionization nuclear emission-line regions (LINERs; 
for a review, see \citealt{2008ARA&A..46..475H}).} with $z \lesssim 0.4$.
For star-forming galaxies, we further apply the following selection criteria:

\begin{enumerate}

\item{To ensure reliable metallicity estimates, a signal-to-noise ratio (S/N) 
$\geq 5$ is required for \OII\ $\lambda3727$, \Hb, \OIII\ $\lambda5007$, \Ha, 
and \NII\ $\lambda6584$.  For a better estimation of the true uncertainties of 
the emission lines, we adopt the scaled uncertainties as suggested by the 
MPA-JHU group.}

\item{To ensure that the metallicity derived from the fiber spectrum 
represents well the global metallicity, the fiber must cover at least 20\% of 
the $g'$-band total photometric light \citep{2005PASP..117..227K}.  We use the 
fiber and the photometric Petrosian $g'$ magnitude to calculate the fiber 
coverage, as in \citet{2008ApJ...681.1183K}.}

\item{Objects with \Ha/\Hb\ $<$ 2.86 are removed. Balmer decrements lower 
than the theoretical value could result from errors in the subtraction of the stellar 
continuum or errors from flux calibration and measurement. 
No S/N cut is set here because the median S/N of observed \Ha/\Hb\ is $\sim$4, and 
the median S/N of \OII\ after extinction correction is already larger than 3.}

\item {Stellar mass estimates are available.}

\end{enumerate}

\noindent
For AGNs, we place the following conditions:

\begin{enumerate}

\item{A S/N $\geq$ 3 is required for \OII, \Hb, \OIII, and \Ha.  The S/N ratio 
cut is more lenient than that of star-forming galaxies because for AGNs we do 
not need to derive the metallicity using the emission lines.  Uncertainties are 
scaled in a similar manner as for the star-forming galaxies.}

\item{A S/N $\geq$ 5 is required for \Ha/\Hb, and objects with \Ha/\Hb\ 
$<$ 3.1 are removed.}

\item {Stellar mass estimates are available.}

\item{The specific SFR (sSFR $\equiv$ SFR/$M_*$) measured in the fiber must 
exceed $10^{-10.5}$ yr$^{-1}$. The fiber SFRs for AGN host galaxies are 
estimated from the relation between the sSFR and the 4000 \AA\ break (D4000) 
established empirically for star-forming galaxies \citep{2004MNRAS.351.1151B}.  
\cite{2016ApJS..227....2S} show that, when sSFR $\gtrsim 10^{-10.5}$ yr$^{-1}$,
D4000-based sSFRs show good agreement with sSFRs derived from global SED 
fitting. For lower values of sSFR, D4000 is less sensitive to 
sSFR, and SFRs derived in this manner are subject to much larger uncertainties. 
The limitations of using D4000 to trace SFRs in AGNs are further 
discussed in Section \ref{sec5.2}.}

\end{enumerate}

The final sample contains 99,355 star-forming galaxies and 5,472 AGNs, both 
groups covering a very similar redshift range of $z \approx 0.02-0.33$. The AGNs
span over 5 orders of magnitude in \OIII\ luminosity ($10^{39.3}-10^{44.4}$ 
erg~s$^{-1}$), from run-of-the-mill type~2 Seyferts to those powerful enough 
to be considered type~2 quasars \citep{2008AJ....136.2373R, 
2018ApJ...859..116K}.  We also include the sample used in \citet[hereinafter 
K04] {2004AJ....127.2002K}, which consists of 97 star-forming galaxies from 
the Nearby Field Galaxies Survey \citep[NFGS;][]{2000ApJS..126..331J, 
2000ApJS..126..271J}. The great increase in the sample size with respect to K04, 
coupled with the homogeneity of the SDSS and ancillary data, motivates us to 
update the \OII\ SFR calibration for star-forming galaxies, as well as to extend it 
to their active counterparts.

\begin{figure*}[t]
\centering
\includegraphics[width=\textwidth]{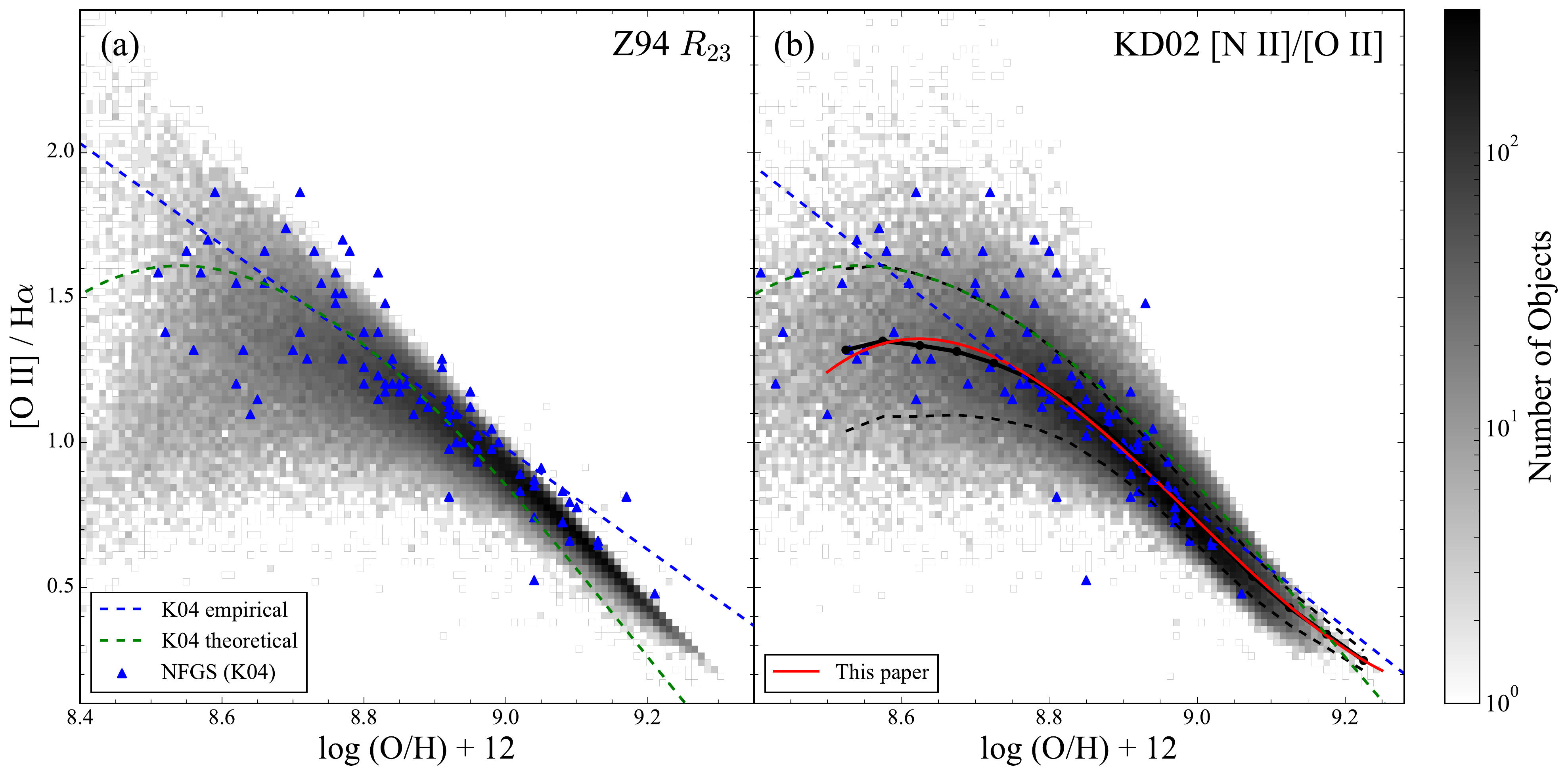}
\caption{The dependence of the extinction-corrected \OII/\Ha\ ratio on oxygen 
abundance $\log \rm{(O/H)} + 12$, calculated from (a) the $R_{23}$ method of 
\citet[][Z94]{1994ApJ...420...87Z} and (b) the \NII/\OII\ method of 
\citet[][KD02]{2002ApJS..142...35K}. Gray squares represent star-forming 
galaxies from SDSS DR7, color-coded by the number of objects. Blue triangles 
are galaxies from the NFGS \citep{2000ApJS..126..331J, 
2000ApJS..126..271J} in K04. The blue dashed line is the empirical fit to the 
NFGS sample, and the green dashed curve is the theoretical 
relation predicted by photoionization models from K04. Black solid and dashed 
curves indicate the median and $\pm 1 \sigma$ \OII/\Ha\ ratio 
for SDSS DR7 star-forming galaxies. The red solid curve is our 
best-fit relation to the data in the range $8.5<\log \rm{(O/H)} + 12<9.2$.  We 
do not fit the data with $\log \rm{(O/H)} + 12<8.5$ because the \NII/\OII\ 
method is less sensitive to abundance in this regime.}
\label{fig1}
\end{figure*}

\begin{figure*}[t]
\centering
\includegraphics[width=\textwidth]{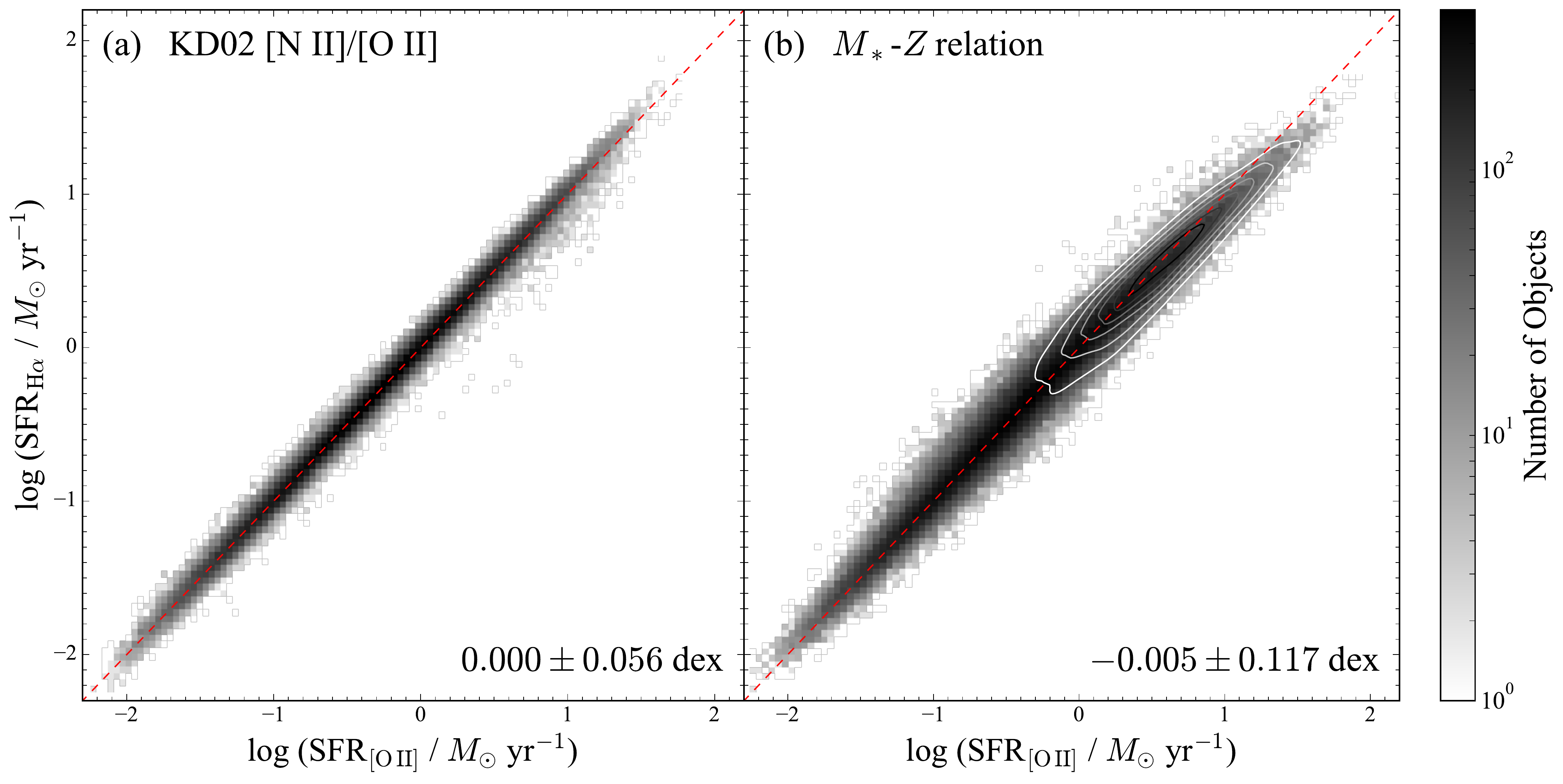}
\caption{Comparison of SFRs from \Ha\ to those from \OII\ with oxygen abundance 
correction from (a) the \NII/\OII\ method and (b) the $M_*$-$Z$ relation using 
\NII/\OII\ method from \citet {2008ApJ...681.1183K}. The median and standard 
deviation of the difference in SFRs (log SFR$_{\rm H\alpha}$ $-$ log 
SFR$_{\rm \OII}$) are shown in the bottom right of each panel. Contours in 
panel (b) indicate the distribution of objects with $M_* > 
10^{10.5}\,M_{\odot}$. Red dashed line denotes 1:1 relation.}
\label{fig2}
\end{figure*}

\section{Calibration of \OII\ SFR indicator in star-forming galaxies} \label{sec3}

\Ha\ provides a robust SFR standard for galaxies at redshifts $\lesssim 0.5$, 
and we use it as the reference to calibrate \OII.  As in K04, we adopt 
Kennicutt's (1998) conversion between extinction-corrected \Ha\ luminosity 
($L_{\rm H\alpha}$) and SFR, after adjustment for our adopted initial mass 
function:

\begin{equation} \label{eq1}
{\rm SFR}(M_{\odot}\ {\rm yr}^{-1}) = {5.3\times10^{-42} L_{\rm H\alpha}\, 
({\rm erg \ s^{-1}})}.
\end{equation}

Apart from dust reddening, metallicity has the strongest effect on the relation
between \Ha\ and \OII\ (K04).  Figure \ref{fig1} shows the relation between the
extinction-corrected \OII/\Ha\ ratio and the gas-phase oxygen 
abundance, $\log \rm{(O/H)} + 12$, estimated indirectly from both the $R_{23}$ 
method\footnote{$R_{23} \equiv$ (\OII\ $\lambda3727$+\OIII\ $\lambda 
\lambda4959, 5007$)/\Hb\ \citep{1979MNRAS.189...95P}.}, as implemented by 
\citet [hereinafter Z94]{1994ApJ...420...87Z}, and the \NII/\OII\ method of 
\citet[hereinafter KD02] {2002ApJS..142...35K}. The $R_{23}$ metallicity 
diagnostic of Z94 is the preferred choice of K04, while \NII/\OII\ 
has low residual discrepancy in relative metallicity, suffers less from AGN 
contamination, and is also less sensitive to the ionization parameter 
\citep{2008ApJ...681.1183K}. K04 provide empirical calibrations for different 
metallicity diagnostics based on observations from NFGS and theoretical 
calibrations from photoionization models.  Both of the methods of deriving 
metallicity used here agree well with the NFGS observations (K04). However, 
on account of the limited size of the NFGS sample, neither the low-metallicity 
nor the high-metallicity end of the population is well covered, and K04's 
calibrations deviate from the sample distribution from SDSS DR7 (Figure 
\ref{fig1}).  This is especially notable for metallicities derived using the 
Z94 $R_{23}$ method (left panel).  Since the KD02 \NII/\OII\ method provides 
better relative metallicities compared to the Z94 $R_{23}$ method 
\citep{2008ApJ...681.1183K}, we adopt the former for the rest of the 
subsequent analysis.  A third-order polynomial fit to the median of the binned 
data yields

\begin{equation} \label{eq2}
\begin{split}
\rm{\OII/H\alpha} = &(-4373.14\pm292.87)+(1463.92\pm98.33)x+ \\
&(-163.045\pm11.0019)x^2+(6.04285\pm0.41019)x^3,
\end{split}
\end{equation}

\noindent
where the oxygen abundance $x \equiv \rm{log (O/H)+12}$.  From KD02,

\begin{equation} \label{eq3}
\begin{split}
\rm{log (O/H)+12} = &\log (1.54020+1.26602 R+ \\
&0.167977 R^2) + 8.93,
\end{split}
\end{equation}

\noindent
with $R = \log \rm{(\NII/\OII)}$.  For objects without coverage or detection 
of \NII, we estimate the metallicity from its relation with stellar mass 
\citep[$M_*$-$Z$ relation;][]{2004ApJ...613..898T}.  The $M_*$-$Z$ relation of 
\citet{2008ApJ...681.1183K}, based on metallicities from the KD02 \NII/\OII\ 
method, is given by

\begin{equation} \label{eq4}
\begin{split}
\rm{log (O/H)+12} = & 28.0974 - 7.23631 \log M_* + \\ 
                                   & 0.850344 (\log M_*)^2 - 0.0318315 (\log M_*)^3.
\end{split}
\end{equation}

\noindent
Although the $M_*$-$Z$ relation of \citet{2008ApJ...681.1183K} is based on 
star-forming galaxies in SDSS DR4, we confirm that it fits our SDSS DR7 sample 
equally well. Since Equation \ref{eq4} only covers the mass range 
$10^{8.5}-10^{11}\,M_{\odot}$, we exclude a small fraction (0.6\%) of the 
objects with $M_* < 10^{8.5}\,M_{\odot}$. For galaxies with  $M_* > 10^{11}\,
M_{\odot}$, we fix the oxygen abundance\footnote{The $M_*$-$Z$ relation 
flattens toward the high-mass end, and hence we use the value of the oxygen 
abundance at $M_* = 10^{11}\,M_{\odot}$.} to $\rm{log (O/H)+12} = 9.02$.

Combining Equations \ref{eq1} and \ref{eq2}, we arrive at 

\begin{equation} \label{eq5}
\begin{split}
&{\rm SFR}(M_{\odot}\ {\rm yr}^{-1}) =  5.3\times10^{-42} L_{\rm \OII} \,({\rm erg \ s^{-1}}) /\\
& (-4373.14+1463.92x -163.045x^2+6.04285x^3).
\end{split}
\end{equation}

\noindent
Figure \ref{fig2} compares the SFRs from \Ha\ (Equation \ref{eq1}) with SFRs 
based on \OII\ (Equation \ref{eq5}), for metallicities estimated using both the 
KD02 \NII/\OII\ method and the $M_*$-$Z$ relation.  No systematic 
differences are found in either panel. For metallicities based on the 
KD02 \NII/\OII\ method, the scatter between the two SFRs is extremely small 
(0.056 dex); the scatter in SFR is higher for the case of metallicities based 
on the $M_*$-$Z$ relation, but it is still small (0.12 dex).  For galaxies more
massive than $M_* \gtrsim 10^{10.5}\, M_{\odot}$ (contours in Figure 
\ref{fig2}b), there is a mild tendency for the \OII-based SFRs to be 
overestimated, but the effect is insignificant (median difference $= 
-0.04 \pm 0.14$ dex).

\OII/\Ha\ can also be affected by ionization parameter ($U$), the 
dimensionless ratio of the incident ionizing photon density to the hydrogen 
density \citep[K04;][]{2006ApJ...642..775M}. \citet{2006ApJ...642..775M} 
found that \OII/\Ha\ depends mildly on $U$. This accounts for the dispersion 
around the median curve of \OII/\Ha\ versus oxygen abundance (Figure 
\ref{fig1}). However, we note that K04's \OII-based SFRs, which are based on 
a theoretical calibration using a single value of $U$ value, agree very well 
with \Ha-based SFRs ($\sim$0.05 dex residual scatter).  Similarly, the 
\OII-based SFRs in this paper, corrected only for oxygen abundance, also 
agree with the \Ha-based SFRs to within 0.056 dex scatter (Figure \ref{fig2}a).
This suggests that the dependence on $U$ at a fixed metallicity must be weak 
and/or star-forming galaxies span a narrow range of $U$. We do not consider 
further the effect of $U$ for star-forming galaxies.

\section{Calibration of \OII\ SFR indicator in active galaxies} \label{sec4}

We perform photoionization calculations using the latest version (C17.01) of 
{\tt CLOUDY} \citep[][]{2017RMxAA..53..385F} to predict the range of allowed 
values of \OII/\OIII\ under realistic physical conditions thought to be 
prevalent in the NLRs.  The models used here are identical to those described 
in \citet{2019ApJ...873..103Z}, who aimed to predict the intensities of \NeII\ 
12.81 \micron\ and \NeIII\ 15.55 \micron\ relative to \NeV\ 14.32 \micron.  As 
\NeV\ can only be excited by AGNs, the restricted range of \NeII/\NeV\ and 
\NeIII/\NeV\ predicted by the photoionization models then implies that the 
observed strength of \NeV\ can be used to subtract the contribution of the NLR 
from the lower-ionization neon lines, which are sensitive to SFR \citep
{2007ApJ...658..314H}. The neon-based SFRs derived for the AGN host galaxies 
by \citet{2019ApJ...873..103Z} agree well with those estimated independently 
from global SED fitting.

For completeness, we briefly describe the setup of our models. We use the 
intrinsic AGN radiation field from \citet{2014MNRAS.438.2253S}, which is based 
on data compilation for a large sample of type~1 AGNs.  We choose three median 
SEDs binned by bolometric luminosity: $\log (L_{\rm bol}$/erg s$^{-1}) >$ 46.3,
45.8$-$46.3, and $<$ 45.8. We construct isobaric, dusty and radiation 
pressure-dominated models by varying the radiation pressure while holding the 
total pressure constant.  With a median $M_* = 10^{10.77}\, M_{\odot}$, the
corresponding oxygen abundance of our AGN sample is $\log \rm{(O/H)} + 12 = 
9.02$, as estimated from the \NII/\OII-based $M_*$-$Z$ relation of \citet
{2008ApJ...681.1183K}. We assume twice solar metallicity \citep[$\log \rm{(O/H)_\odot}
 + 12 = 8.69$;][]{2001ApJ...556L..63A} and the dust composition and size 
distribution of the Orion nebula. The absolute amount of dust is scaled to 
twice the Orion value. We vary the ionization parameter $\log U = -3$ to $2$ 
in steps of 0.5, and we adjust the initial hydrogen 
density\footnote{The initial hydrogen density is the density at the 
illuminated surface of a cloud, which does not necessarily equal 
the density of where emission lines arise. In a constant-density model, the 
gas density stays constant from the illuminated surface to the ionization 
front. While in our isobaric model, when radiation pressure is 
dominant, the gas density increases as the radiation goes deeper into the 
cloud. See Figure 3 in \citet {2014MNRAS.438..901S} for more details.} 
from $n_{\rm H} = 1$ to $10^3$ cm$^{-3}$, in steps of 1 in the logarithm. 

Radiation pressure-dominated NLR models provide a successful framework for 
interpreting spatially resolved spectroscopic observations of nearby Seyfert 
galaxies.  \citet{2014MNRAS.438..901S} showed that when radiation pressure 
dominates the total pressure, the effective ionization parameter is $\sim$0.03,
and the final gas density $n \propto r^{-2}$ (see Equation \ref{eq8}). A 
radially stratified NLR arises from the tendency for forbidden transitions to 
emit most efficiently at densities close to their respective critical 
densities for collisional de-excitation ($n_{\rm crit}$).
For instance, \OIII\ $\lambda5007$, with $n_{\rm crit}=10^{5.8}$ cm$^{-3}$, 
reaches its peak emissivity at a radius $\sim$20 times smaller than that of 
\OII\ $\lambda3727$, which has $n_{\rm crit} \approx 10^{3}$ cm$^{-3}$. 

\begin{figure}[t]
\centering
\includegraphics[width=0.48\textwidth]{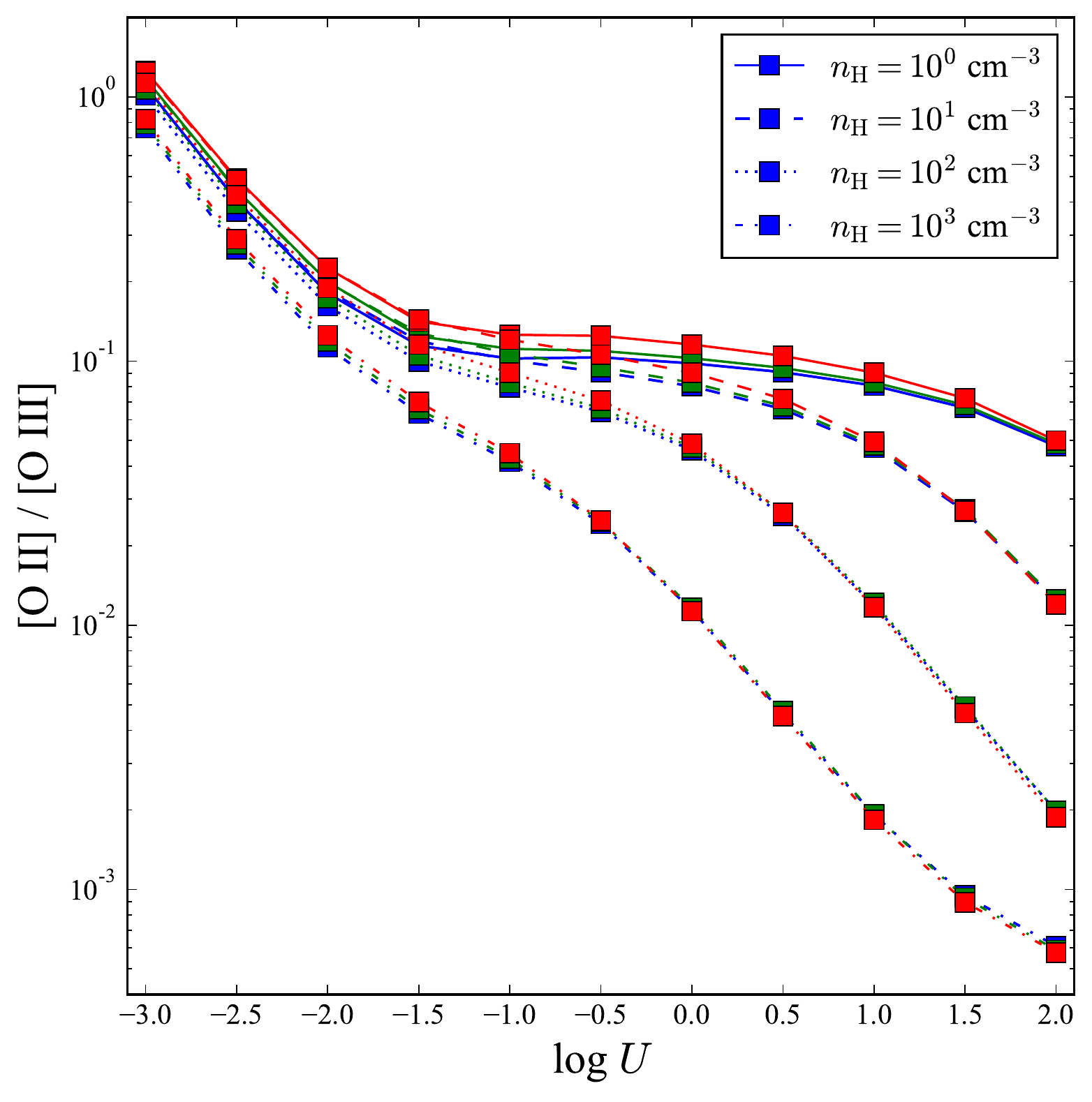}
\caption{The dependence of \OII/\OIII\ on ionization parameter $U$, for initial 
hydrogen particle densities from $n_{\rm H} = 1$ to $10^3$ cm$^{-3}$.  We 
use three input SEDs: $\log (L_{\rm bol}$/erg s$^{-1})$ $>$ 46.3 (red), 
45.8$-$46.3 (green), and $<$ 45.8 (blue).}
\label{fig3}
\end{figure}

Figure \ref{fig3} illustrates the variation of \OII/\OIII\ as a function of $U$
and $n_{\rm H}$.  \OII/\OIII\ decreases with increasing $U$ for $\log U 
\lesssim -1.5$, the regime in which the gas pressure dominates the total 
pressure. In this situation, the gas density remains nearly constant from the 
illuminated surface of the cloud to the ionization front 
\citep{2014MNRAS.438..901S}. For $\log U \gg -1.5$, radiation pressure 
dominates the total pressure. From the definition of $U$, increasing $U$ at 
fixed $n_{\rm H}$ or increasing $n_{\rm H}$ at fixed $U$ will have the same 
effect as increasing the ionizing photon density. At the same ionizing photon 
density (constant product of $U$ and $n_{\rm H}$), \OII/\OIII\ is almost a 
constant.  Increasing the ionizing photon density is equivalent to putting a 
cloud closer to the central ionizing source. Therefore, \OII/\OIII\ decreases 
with both higher $U$ ($\log U \gtrsim 0$ at $n_{\rm H} = 1 $ cm$^{-3}$) and 
higher $n_{\rm H}$, due to collisional de-excitation of \OII\ at densities 
above its critical density. Choosing the \OII/\OIII\ ratio at $\log U = -0.5$ 
and $n_{\rm H} = 1$ cm$^{-3}$ (to ensure that the cloud is radiation 
pressure-dominated but \OII\ not yet collisionally de-excited), we obtain 

\begin{equation} \label{eq6}
L_{\rm \OII}= 0.109^{+0.016}_{-0.006}\, L_{\rm \OIII},
\end{equation}

\noindent
where $L_{\rm \OII}$ and $L_{\rm \OIII}$ are purely from the NLR, which we 
assume to be radiation pressure-dominated, and the uncertainties reflect the 
different input AGN SEDs. We only use the ratio where both \OII\ and \OIII\ 
reach their maximum emissivity, which means that the density is below 
$n_{\rm crit}$ of \OII. 
The true global value of \OII/\OIII\ is likely slightly lower, considering the 
contribution from the inner part of the NLR, where the density is probably 
higher than $n_{\rm crit}$ of \OII\ and hence would produce lower \OII/\OIII. 
Determining an accurate global value of \OII/\OIII\ depends on the radial 
variation of the covering factor. A reasonable assumption of constant covering 
factor over radius suggests that the total emission is dominated by emission 
where $n\approx n_{\rm crit}$ \citep{2014MNRAS.438..901S}. Hence, our 
predicted values here should be good estimates.

We assume that all of the \OIII\ emission arises from the AGN, a reasonable 
supposition for AGN hosts, which are generally massive (metal-rich) galaxies 
\citep{2005ApJ...629..680H}.  Combining Equations \ref{eq5} and \ref{eq6} yields

\begin{equation} \label{eq7}
\begin{split}
&{\rm SFR}(M_{\odot}\ {\rm yr}^{-1}) = 5.3\times10^{-42} (L_{\rm \OII}-0.109L_{\rm \OIII}) ({\rm erg\ s^{-1}})/\\
&(-4373.14+1463.92x -163.045x^2+6.04285x^3),
\end{split}
\end{equation}

\noindent
where $L_{\rm \OII}$ and $L_{\rm \OIII}$ are the total, extinction-corrected
\OII\ and \OIII\ luminosities.

Strictly speaking, \HII\ regions, of course, also
emit \OIII, but \OII/\OIII\ increases with increasing stellar mass and 
increasing metallicity \citep{2014MNRAS.442..900N}. In our star-forming galaxy 
sample, the distribution of \OII/\OIII\ flattens and approaches a median value 
of $\sim$6 when $M_* > 10^{10}\, M_{\odot}$.  As AGNs predominantly reside in 
massive systems \citep{2003ApJ...583..159H, 2003MNRAS.346.1055K}, after 
accounting for \OIII\ emission produced by star formation with $\rm \OII/\OIII=6$, 
we find that the NLR contributes 88.5\% of the total \OIII\ emission and 12.2\% of 
total \OII\ emission in our AGN sample.  Neglecting the \OIII\ produced by star 
formation will induce a median difference of only 1.6\% on \OII\ (increased to 
13.8\%), which is smaller than the uncertainties introduced by using the $M_*$-$Z$ 
relation to estimate the metallicity.

\begin{figure*}[t]
\centering
\includegraphics[width=\textwidth]{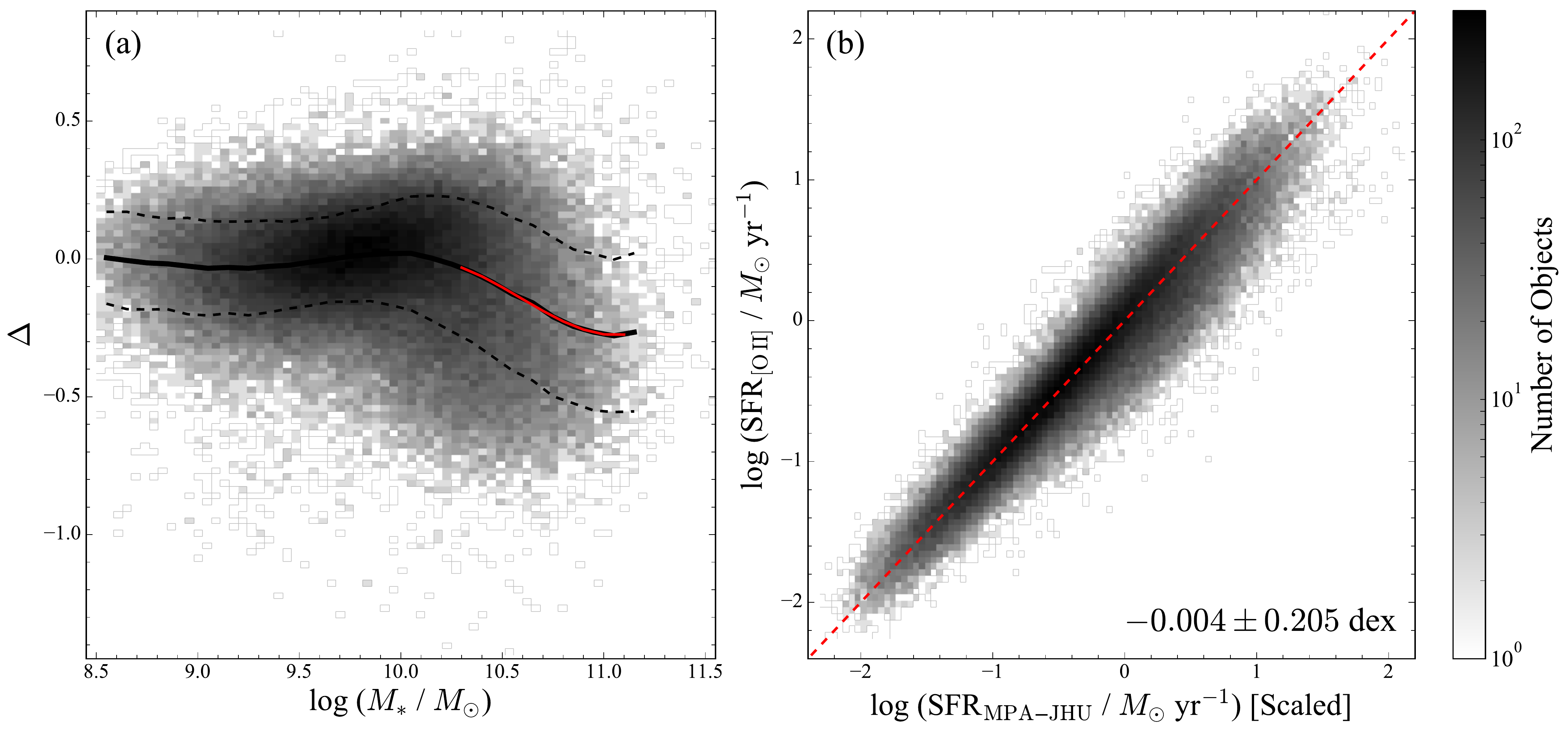}
\caption{(a) The difference between SFRs from \OII\ and those provided by the 
MPA-JHU catalog versus stellar mass for star-forming galaxies. Solid and dashed black 
curves represent the median and $\pm 1\sigma$ of the data. The red solid 
curve is our fit to the objects with $M_*=10^{10.3}-10^{11.1}\,M_{\odot}$:
$\Delta \equiv \log({\rm SFR_{\OII}}) - \log({\rm SFR_{MPA{\text-}JHU}}) = -786.728 + 
223.720\log M_* - 21.1686(\log M_*)^2 + 0.666370(\log M_*)^3$.  (b) Comparison of SFRs 
from \OII\ to catalog values (after scaling). The median and standard deviation
of the difference are shown on the bottom-right corner of the panel.}
\label{fig4}
\end{figure*}

\section{Discussion} \label{sec5}

\subsection{Intrinsic $\rm{\OII/\OIII}$ Ratio in AGNs} \label{sec5.1}

Our calculations indicate that over a wide range of plausible conditions the 
NLR of high-ionization \citep[highly accreting;][]{2009ApJ...699..626H} AGNs 
emits a highly restricted ratio of \OII/\OIII\ $\approx 0.1$ (Figure \ref{fig3}), 
and the resulting median fraction of \OII\ from AGNs in our 
sample is 13.8\%.  These values are somewhat smaller than found by 
other investigators.  For example, \citet{2006ApJ...642..702K} find 
\OII/\OIII\ ratios 0.03$-$4, with a typical value of $\sim$0.3.  \citet
{2014MNRAS.444.3961D} quote an AGN contribution to \OII\ of $\sim$40\% 
for three nearby AGNs, and the sample of \citet{2018ApJ...861L...2T} 
indicates AGN fractions of $\sim$20\% to 100\%.
This discrepancy can be traced to the smaller ionization parameters invoked 
in  previous calculations ($\log U \ll -1.5$), which effectively produce 
constant-density models instead of radiation pressure-dominated models 
(Section \ref{sec4}). When radiation pressure does not dominate the total 
pressure, the local ionization parameter can be smaller than $\log U \approx 
-1.5$, which produces larger \OII/\OIII.  Here we show that the typical sizes 
of the radiation pressure-dominated region in the NLR are large and comparable 
to the dimensions directly observed in well-studied sources.  

If the radiation pressure dominates the total gas pressure, the 
radius of the outer boundary of the NLR can be approximated by 
\citep[their Equation 6] {2014MNRAS.438..901S}

\begin{equation} \label{eq8}
r = 1.4 \times 10^4 L^{0.5}_{\rm ion, 45} n^{-0.5} T_4^{-0.5} \, {\rm pc},
\end{equation}

\noindent
where $n$ is the gas density in units of cm$^{-3}$, $T_4$ is the 
temperature in units of $10^4$ K near the ionization front, and 
$L_{\rm ion, 45}$ is the ionizing luminosity at 1--1000 Ryd in units of 
$10^{45}$ erg s$^{-1}$.  We use the extinction-corrected 
\OIII\ luminosity to estimate the AGN bolometric luminosity, adopting the
luminosity-dependent bolometric corrections from \citet{2009A&A...504...73L}, 
as parameterized by \citet{2015ApJ...811...26T}. We infer from the three input 
SEDs used in our models $L_{\rm ion}/L_{\rm bol} = 0.31$ for 
$\log (L_{\rm bol}$/erg s$^{-1}) \leq 46.3$ and $L_{\rm ion}/L_{\rm bol} = 
0.25$ for $\log (L_{\rm bol}$/erg s$^{-1}) > 46.3$. As in 
\citet{2014MNRAS.438..901S}, we assume a typical interstellar medium pressure 
of $nT_4 = 0.3$ cm$^{-3}$ \citep{2011piim.book.....D}. 
For the median luminosity of our sample, $\log (L_{\rm \OIII}$/erg s$^{-1}) = 
41.4$, radiation pressure dominates at a radius of $r \approx 3.3$ kpc.  This 
estimate compares favorably with observations.  Spatially resolved studies 
have established an empirical relation between the physical extent and the 
luminosity of the NLR \citep[e.g.,][]{2002ApJ...574L.105B, 2019MNRAS.486.5075N}.
From the radius-luminosity relation of \citet{2011ApJ...732....9G}, the 
median $L_{\rm \OIII}$ of our sample corresponds to $r = 4.2 \pm 0.8$ kpc, 
in reasonably good agreement with the predicted size of radiation 
pressure-dominated NLRs.

\begin{figure*}[t]
\centering
\includegraphics[width=\textwidth]{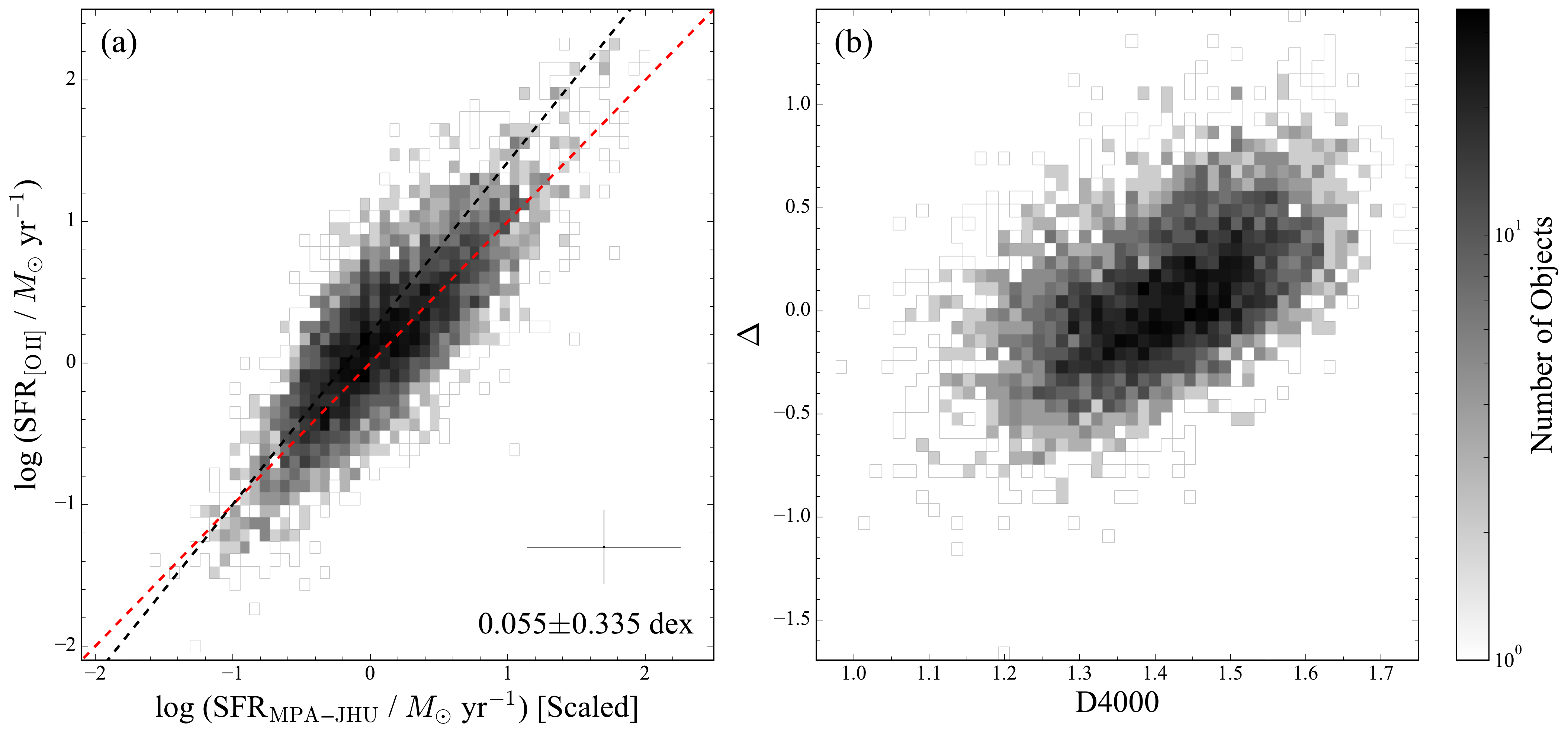}
\caption{(a) Comparison of SFRs derived from \OII\ with those from the MPA-JHU 
catalog for AGNs, with AGN contribution subtracted. The MPA-JHU SFRs are 
based on the relation between sSFR and D4000 in star-forming galaxies \citep
{2004MNRAS.351.1151B}, with SFRs scaled in the same manner as for 
the star-forming galaxies (Figure \ref{fig4}). The median and standard 
deviation of the difference, as well as typical errors for both SFRs, are 
shown on the bottom-right corner of the panel.  Red and black 
dashed lines denote the 1:1 relation and the least squares fit to the data, 
respectively. (b) The difference between SFRs from \OII\ and 
those provided by the MPA-JHU catalog (scaled) versus D4000 for AGNs.
The squares are color-coded by the number of objects.  }
\label{fig5}
\end{figure*}

\subsection{External Comparison of SFRs} \label{sec5.2}

In order to show quantitatively the robustness of our new \OII-based SFR 
estimator for AGNs, we compare our SFRs with independent values derived from 
the D4000 method, which depend solely on the properties of the stellar 
population in the host galaxy. \citet{2004MNRAS.351.1151B} calibrated the 
relation between sSFR and D4000 in star-forming galaxies and applied it to 
AGNs.  It is important to acknowledge that AGNs may introduce additional 
uncertainties.   Nearby AGNs predominantly live in massive, evolved systems 
\citep[e.g.,][]{2003ApJ...583..159H, 2003MNRAS.341...33K}, a regime where 
the sSFR-D4000 relation suffers the largest dispersion.  Moreover, scattered light 
from the nonstellar nucleus \citep[e.g.,][]{1985ApJ...297..621A, 2016MNRAS.456.2861O} 
contributes featureless continuum to the integrated spectrum, diluting
D4000.  Notwithstanding these complications, D4000 offers the only practical
avenue to estimate SFRs for the SDSS AGN sample under consideration.  We 
perform this external check in two steps, first by verifying whether our 
updated \OII-based estimator yields SFRs consistent with the \Ha-based SFRs 
from the MPA-JHU catalog for star-forming galaxies, and then repeating the 
same cross-check for the active galaxies.

Our SFRs, following \citet{1998ARA&A..36..189K}, assume that no Lyman continuum 
photons are absorbed by dust.  This is a reasonable approximation, validated 
by the overall agreement between SFRs derived from \Ha\ and far-infrared 
continuum \citep{2006ApJ...642..775M}.  Meanwhile, the MPA-JHU SFRs for 
star-forming galaxies, based on \Ha, account for absorption of ionizing photons
by dust within the \HII\ regions \citep{2001MNRAS.323..887C, 
2002MNRAS.330..876C}.  In order to compare our \OII-based SFRs to those 
provided by the MPA-JHU catalog for AGNs, we must first ensure that our 
\OII-based SFRs for star-forming galaxies agree with their corresponding 
MPA-JHU SFRs.  Recall, however, that the \OII-based SFRs involve a metallicity 
term.  Whereas in star-forming galaxies the oxygen abundance can be inferred 
from \NII/\OII\ (Equation \ref{eq3}), this option is not available for AGNs 
because of contamination from the NLR, and we must resort to the $M_*$-$Z$ 
relation (Equation \ref{eq4}).  To mimic as closely as possible the situation 
for AGNs, for this comparison our \OII-based SFRs for star-forming galaxies 
{\it also}\ adopt oxygen abundances estimated from the $M_*$-$Z$ relation.  
We use stellar masses from the MPA-JHU catalog.

Figure \ref{fig4}a compares the two sets of SFRs as a function of stellar 
mass for star-forming galaxies. Systematic residuals are clearly present, and 
they depend on stellar mass.  
While $\Delta \equiv \log({\rm SFR_{\OII}}) - 
\log({\rm SFR_{MPA{\text-}JHU}}) \approx -0.007 \pm 0.184$ dex for $M_* 
\lesssim 10^{10.3}\,M_{\odot}$, at higher masses the residuals become large 
and mainly negative ($\Delta = -0.125 \pm 0.277$ dex), reminiscent of the 
trend found in \citet[their Figure 8]{2004MNRAS.351.1151B}.  While plausible 
explanations have been offered for these trends \citep{2001MNRAS.323..887C, 
2002MNRAS.330..876C, 2004MNRAS.351.1151B}, of practical relevance here is that 
the systematic trend at high masses($M_* = 
10^{10.3}-10^{11.1}\,M_{\odot}$) can be removed with a third-order polynomial 
(Figure \ref{fig4}, red curve).  
This allows us to compare SFRs from the MPA-JHU catalog with 
those derived from our \OII\ method for AGNs. 
We do not adjust the objects with  $M_* < 10^{10.3}\, M_{\odot}$, 
and for objects with $M_* > 10^{11.1}\,M_{\odot}$ we set $\Delta = -0.274$ dex, 
the median difference at $M_* = 10^{11.1}\, M_{\odot}$.
After applying these corrections, the MPA-JHU SFRs are brought into good 
agreement with our \OII-based SFRs (Figure \ref{fig4}b), 
with the median difference for objects with $M_* > 10^{10.3}\, 
M_{\odot}$ brought from $-0.125$ dex to $0.000$ dex. 

Figure \ref{fig5}a compares the SFRs from our \OII\ calibration for AGNs 
(Equation \ref{eq7}) with SFRs based on the D4000 method from the MPA-JHU 
catalog.  For the sample as a whole, $\Delta = 0.055 \pm 0.335$ dex, which we 
regard as quite satisfactory agreement, in light of the substantial 
uncertainties ($\sim$0.6 dex) inherent in the crude D4000-based SFRs. 
In detail, however, it is apparent that the two sets of 
measurements exhibit a slight departure from a 1:1 relation.  A formal least 
squares fit yields a slope of 1.21 (black dashed line).  Moreover, the residual
difference between the \OII-based SFRs and MPA-JHU SFRs varies systematically 
with D4000 (Figure \ref{fig5}b).   This systematic trend, which is unphysical,
originates from the fact that the MPA-JHU SFRs were estimated using a 
relation between sSFR and D4000 originally derived for star-forming galaxies 
\citep{2004MNRAS.351.1151B}. We verify that most of the AGNs with 
positive residuals (positive $\Delta$) and large D4000 tend to be located below 
the star-forming galaxy main sequence \citep{2007ApJ...660L..43N}.  The 
sSFR-D4000 relation is particularly unreliable for estimating SFRs of galaxies 
in this regime.  This, coupled with the fact that \OII\ probes SFRs on shorter 
timescales than D4000, may be responsible for the residual discrepancies observed 
in Figure \ref{fig5}b, which we do not consider to be serious.

\section{Summary} \label{sec6}

We use an extensive sample of star-forming galaxies and narrow-line AGNs from 
SDSS DR7 to investigate the use of \OII\ $\lambda 3727$ as a SFR indicator. The 
large sample enables us to probe a wide dynamic range in physical properties 
of galaxies, spanning $\sim$4 dex in SFR, more than 3 dex in stellar mass, 
and nearly 1 dex in metallicity.

Our main results are as follows:

\begin{itemize}
{\item Consistent with previous work, we show that the \OII/\Ha\ ratio in 
star-forming galaxies depends strongly on metallicity.  Using the better 
understood \Ha\ SFR indicator as reference, we expand upon the work 
of \citet{2004AJ....127.2002K} and propose a new empirical SFR calibration 
based on \OII\ that explicitly includes a correction for metallicity estimated 
from the \NII/\OII\ ratio.}

{\item Our \OII-based SFRs for star-forming galaxies show excellent 
consistency with SFRs obtained from \Ha; the scatter is 0.056 dex for 
metallicities derived from the \NII/\OII\ method, increasingly only to 
0.12 dex for less accurate metallicities inferred from the mass-metallicity 
relation.}

{\item With the aid of a set of photoionization models designed to mimic the 
conditions of the NLR, we demonstrate that high-ionization AGNs (e.g., Seyfert 
galaxies and quasars) emit a remarkably constant ratio of \OII\ 
$\lambda 3727$/\OIII\ $\lambda 5007$.  We introduce a new formalism for 
estimating SFRs in AGNs based on \OII. Our methodology assumes that all of 
the \OIII\ emission arises from the NLR, a reasonable approximation for massive 
host galaxies.}

{\item The \OII-based SFRs agree reasonably well with independent SFRs derived 
from the stellar continuum (D4000) of the AGN host galaxies, demonstrating the 
robustness and effectiveness of our new method.}

\end{itemize}

\acknowledgments 
We are grateful to an anonymous referee for constructive comments and suggestions. 
We thank Jinyi Shangguan for helpful discussions.  This work was supported by the 
National Key R\&D Program of China (2016YFA0400702) and the National Science 
Foundation of China (11721303).

\vspace{5mm}
\software{Astropy \citep{2013A&A...558A..33A, 2018AJ....156..123A},  
Cloudy \citep{2017RMxAA..53..385F},
Matplotlib \citep{Hunter:2007},
Numpy \citep{numpy},
Scipy \citep{scipy}
}

%% To help institutions obtain information on the effectiveness of their 
%% telescopes the AAS Journals has created a group of keywords for telescope 
%% facilities.
%
%% Following the acknowledgments section, use the following syntax and the
%% \facility{} or \facilities{} macros to list the keywords of facilities used 
%% in the research for the paper.  Each keyword is check against the master 
%% list during copy editing.  Individual instruments can be provided in 
%% parentheses, after the keyword, but they are not verified.

%\vspace{5mm}
%\facilities{HST(STIS), Swift(XRT and UVOT), AAVSO, CTIO:1.3m,
%CTIO:1.5m,CXO}

%% Similar to \facility{}, there is the optional \software command to allow 
%% authors a place to specify which programs were used during the creation of 
%% the manusscript. Authors should list each code and include either a
%% citation or url to the code inside ()s when available.

%\software{astropy \citep{2013A&A...558A..33A},  
%          Cloudy \citep{2013RMxAA..49..137F}, 
%          SExtractor \citep{1996A&AS..117..393B}
%          }

%\appendix


\begin{thebibliography}{}
\expandafter\ifx\csname natexlab\endcsname\relax\def\natexlab#1{#1}\fi

\bibitem[{{Abazajian} {et~al.}(2009){Abazajian}, {Adelman-McCarthy},
  {Ag{\"u}eros}, {Allam}, {Allende Prieto}, {An}, {Anderson}, {Anderson},
  {Annis}, {Bahcall}, \& et~al.}]{2009ApJS..182..543A}
{Abazajian}, K.~N., {Adelman-McCarthy}, J.~K., {Ag{\"u}eros}, M.~A., {et~al.}
  2009, \apjs, 182, 543

\bibitem[{{Allende Prieto} {et~al.}(2001){Allende Prieto}, {Lambert}, \&
  {Asplund}}]{2001ApJ...556L..63A}
{Allende Prieto}, C., {Lambert}, D.~L., \& {Asplund}, M. 2001, \apjl, 556, L63

\bibitem[{{Antonucci} \& {Miller}(1985)}]{1985ApJ...297..621A}
{Antonucci}, R.~R.~J., \& {Miller}, J.~S. 1985, \apj, 297, 621

\bibitem[{{Argence} \& {Lamareille}(2009)}]{2009A&A...495..759A}
{Argence}, B., \& {Lamareille}, F. 2009, \aap, 495, 759

\bibitem[{{Astropy Collaboration} {et~al.}(2013){Astropy Collaboration},
  {Robitaille}, {Tollerud}, {Greenfield}, {Droettboom}, {Bray}, {Aldcroft},
  {Davis}, {Ginsburg}, {Price-Whelan}, {Kerzendorf}, {Conley}, {Crighton},
  {Barbary}, {Muna}, {Ferguson}, {Grollier}, {Parikh}, {Nair}, {Unther},
  {Deil}, {Woillez}, {Conseil}, {Kramer}, {Turner}, {Singer}, {Fox}, {Weaver},
  {Zabalza}, {Edwards}, {Azalee Bostroem}, {Burke}, {Casey}, {Crawford},
  {Dencheva}, {Ely}, {Jenness}, {Labrie}, {Lim}, {Pierfederici}, {Pontzen},
  {Ptak}, {Refsdal}, {Servillat}, \& {Streicher}}]{2013A&A...558A..33A}
{Astropy Collaboration}, {Robitaille}, T.~P., {Tollerud}, E.~J., {et~al.} 2013,
  \aap, 558, A33

\bibitem[{{Astropy Collaboration} {et~al.}(2018){Astropy Collaboration},
  {Price-Whelan}, {Sip{\H{o}}cz}, {G{\"u}nther}, {Lim}, {Crawford}, {Conseil},
  {Shupe}, {Craig}, \& {Dencheva}}]{2018AJ....156..123A}
{Astropy Collaboration}, {Price-Whelan}, A.~M., {Sip{\H{o}}cz}, B.~M., {et~al.}
  2018, \aj, 156, 123

\bibitem[{{Baldwin} {et~al.}(1981){Baldwin}, {Phillips}, \&
  {Terlevich}}]{1981PASP...93....5B}
{Baldwin}, J.~A., {Phillips}, M.~M., \& {Terlevich}, R. 1981, \pasp, 93, 5

\bibitem[{{Bennert} {et~al.}(2002){Bennert}, {Falcke}, {Schulz}, {Wilson}, \&
  {Wills}}]{2002ApJ...574L.105B}
{Bennert}, N., {Falcke}, H., {Schulz}, H., {Wilson}, A.~S., \& {Wills}, B.~J.
  2002, \apj, 574, L105

\bibitem[{{Brinchmann} {et~al.}(2004){Brinchmann}, {Charlot}, {White},
  {Tremonti}, {Kauffmann}, {Heckman}, \& {Brinkmann}}]{2004MNRAS.351.1151B}
{Brinchmann}, J., {Charlot}, S., {White}, S.~D.~M., {et~al.} 2004, \mnras, 351,
  1151

\bibitem[{{Cardelli} {et~al.}(1989){Cardelli}, {Clayton}, \&
  {Mathis}}]{1989ApJ...345..245C}
{Cardelli}, J.~A., {Clayton}, G.~C., \& {Mathis}, J.~S. 1989, \apj, 345, 245

\bibitem[{{Chabrier}(2003)}]{2003PASP..115..763C}
{Chabrier}, G. 2003, \pasp, 115, 763

\bibitem[{{Charlot} {et~al.}(2002){Charlot}, {Kauffmann}, {Longhetti},
  {Tresse}, {White}, {Maddox}, \& {Fall}}]{2002MNRAS.330..876C}
{Charlot}, S., {Kauffmann}, G., {Longhetti}, M., {et~al.} 2002, \mnras, 330,
  876

\bibitem[{{Charlot} \& {Longhetti}(2001)}]{2001MNRAS.323..887C}
{Charlot}, S., \& {Longhetti}, M. 2001, \mnras, 323, 887

\bibitem[{{Davies} {et~al.}(2014){Davies}, {Kewley}, {Ho}, \&
  {Dopita}}]{2014MNRAS.444.3961D}
{Davies}, R.~L., {Kewley}, L.~J., {Ho}, I.~T., \& {Dopita}, M.~A. 2014, \mnras,
  444, 3961

\bibitem[{{Draine}(2011)}]{2011piim.book.....D}
{Draine}, B.~T. 2011, {Physics of the Interstellar and Intergalactic Medium}

\bibitem[{{Ferland} {et~al.}(2017){Ferland}, {Chatzikos}, {Guzm{\'a}n},
  {Lykins}, {van Hoof}, {Williams}, {Abel}, {Badnell}, {Keenan}, {Porter}, \&
  {Stancil}}]{2017RMxAA..53..385F}
{Ferland}, G.~J., {Chatzikos}, M., {Guzm{\'a}n}, F., {et~al.} 2017, \rmxaa, 53,
  385

\bibitem[{{Ferrarese} \& {Merritt}(2000)}]{2000ApJ...539L...9F}
{Ferrarese}, L., \& {Merritt}, D. 2000, \apj, 539, L9

\bibitem[{{Gallagher} {et~al.}(1989){Gallagher}, {Bushouse}, \&
  {Hunter}}]{1989AJ.....97..700G}
{Gallagher}, J.~S., {Bushouse}, H., \& {Hunter}, D.~A. 1989, \aj, 97, 700

\bibitem[{{Gebhardt} {et~al.}(2000){Gebhardt}, {Bender}, {Bower}, {Dressler},
  {Faber}, {Filippenko}, {Green}, {Grillmair}, {Ho}, {Kormendy}, {Lauer},
  {Magorrian}, {Pinkney}, {Richstone}, \& {Tremaine}}]{2000ApJ...539L..13G}
{Gebhardt}, K., {Bender}, R., {Bower}, G., {et~al.} 2000, \apj, 539, L13

\bibitem[{{Gilbank} {et~al.}(2010){Gilbank}, {Baldry}, {Balogh}, {Glazebrook},
  \& {Bower}}]{2010MNRAS.405.2594G}
{Gilbank}, D.~G., {Baldry}, I.~K., {Balogh}, M.~L., {Glazebrook}, K., \&
  {Bower}, R.~G. 2010, \mnras, 405, 2594

\bibitem[{{Greene} {et~al.}(2011){Greene}, {Zakamska}, {Ho}, \&
  {Barth}}]{2011ApJ...732....9G}
{Greene}, J.~E., {Zakamska}, N.~L., {Ho}, L.~C., \& {Barth}, A.~J. 2011, \apj,
  732, 9

\bibitem[{{Heckman} \& {Best}(2014)}]{2014ARA&A..52..589H}
{Heckman}, T.~M., \& {Best}, P.~N. 2014, Annual Review of Astronomy and
  Astrophysics, 52, 589

\bibitem[{{Ho}(2004)}]{2004coa..book.....H}
{Ho}, L. 2004, {Carnegie Observatories Astrophysics 4 Volume Hardback Set}
  (Cambridge University Press)

\bibitem[{{Ho}(2005)}]{2005ApJ...629..680H}
{Ho}, L.~C. 2005, \apj, 629, 680

\bibitem[{{Ho}(2008)}]{2008ARA&A..46..475H}
---. 2008, \araa, 46, 475

\bibitem[{{Ho}(2009)}]{2009ApJ...699..626H}
---. 2009, \apj, 699, 626

\bibitem[{{Ho} {et~al.}(2003){Ho}, {Filippenko}, \&
  {Sargent}}]{2003ApJ...583..159H}
{Ho}, L.~C., {Filippenko}, A.~V., \& {Sargent}, W.~L.~W. 2003, \apj, 583, 159

\bibitem[{{Ho} \& {Keto}(2007)}]{2007ApJ...658..314H}
{Ho}, L.~C., \& {Keto}, E. 2007, \apj, 658, 314

\bibitem[{Hunter(2007)}]{Hunter:2007}
Hunter, J.~D. 2007, Computing in Science \& Engineering, 9, 90

\bibitem[{{Jansen} {et~al.}(2000{\natexlab{a}}){Jansen}, {Fabricant}, {Franx},
  \& {Caldwell}}]{2000ApJS..126..331J}
{Jansen}, R.~A., {Fabricant}, D., {Franx}, M., \& {Caldwell}, N.
  2000{\natexlab{a}}, \apjs, 126, 331

\bibitem[{{Jansen} {et~al.}(2000{\natexlab{b}}){Jansen}, {Franx}, {Fabricant},
  \& {Caldwell}}]{2000ApJS..126..271J}
{Jansen}, R.~A., {Franx}, M., {Fabricant}, D., \& {Caldwell}, N.
  2000{\natexlab{b}}, \apjs, 126, 271

\bibitem[{Jones {et~al.}(2001--)Jones, Oliphant, Peterson, {et~al.}}]{scipy}
Jones, E., Oliphant, T., Peterson, P., {et~al.} 2001--, {SciPy}: Open source
  scientific tools for {Python}, , , [Online; accessed 2019-05-27]

\bibitem[{{Kauffmann} {et~al.}(2003{\natexlab{a}}){Kauffmann}, {Heckman},
  {White}, {Charlot}, {Tremonti}, {Brinchmann}, {Bruzual}, {Peng}, {Seibert},
  {Bernardi}, {Blanton}, {Brinkmann}, {Castander}, {Cs{\'a}bai}, {Fukugita},
  {Ivezic}, {Munn}, {Nichol}, {Padmanabhan}, {Thakar}, {Weinberg}, \&
  {York}}]{2003MNRAS.341...33K}
{Kauffmann}, G., {Heckman}, T.~M., {White}, S.~D.~M., {et~al.}
  2003{\natexlab{a}}, \mnras, 341, 33

\bibitem[{{Kauffmann} {et~al.}(2003{\natexlab{b}}){Kauffmann}, {Heckman},
  {Tremonti}, {Brinchmann}, {Charlot}, {White}, {Ridgway}, {Brinkmann},
  {Fukugita}, {Hall}, {Ivezi{\'c}}, {Richards}, \&
  {Schneider}}]{2003MNRAS.346.1055K}
{Kauffmann}, G., {Heckman}, T.~M., {Tremonti}, C., {et~al.} 2003{\natexlab{b}},
  \mnras, 346, 1055

\bibitem[{{Kennicutt} \& {Evans}(2012)}]{2012ARA&A..50..531K}
{Kennicutt}, R.~C., \& {Evans}, N.~J. 2012, \araa, 50, 531

\bibitem[{{Kennicutt}(1992)}]{1992ApJ...388..310K}
{Kennicutt}, Jr., R.~C. 1992, \apj, 388, 310

\bibitem[{{Kennicutt}(1998)}]{1998ARA&A..36..189K}
---. 1998, \araa, 36, 189

\bibitem[{{Kennicutt} {et~al.}(2009){Kennicutt}, {Hao}, {Calzetti},
  {Moustakas}, {Dale}, {Bendo}, {Engelbracht}, {Johnson}, \&
  {Lee}}]{2009ApJ...703.1672K}
{Kennicutt}, Jr., R.~C., {Hao}, C.-N., {Calzetti}, D., {et~al.} 2009, \apj,
  703, 1672

\bibitem[{{Kewley} \& {Dopita}(2002)}]{2002ApJS..142...35K}
{Kewley}, L.~J., \& {Dopita}, M.~A. 2002, \apjs, 142, 35

\bibitem[{{Kewley} \& {Ellison}(2008)}]{2008ApJ...681.1183K}
{Kewley}, L.~J., \& {Ellison}, S.~L. 2008, \apj, 681, 1183

\bibitem[{{Kewley} {et~al.}(2004){Kewley}, {Geller}, \&
  {Jansen}}]{2004AJ....127.2002K}
{Kewley}, L.~J., {Geller}, M.~J., \& {Jansen}, R.~A. 2004, \aj, 127, 2002

\bibitem[{{Kewley} {et~al.}(2005){Kewley}, {Jansen}, \&
  {Geller}}]{2005PASP..117..227K}
{Kewley}, L.~J., {Jansen}, R.~A., \& {Geller}, M.~J. 2005, \pasp, 117, 227

\bibitem[{{Kim} {et~al.}(2006){Kim}, {Ho}, \& {Im}}]{2006ApJ...642..702K}
{Kim}, M., {Ho}, L.~C., \& {Im}, M. 2006, \apj, 642, 702

\bibitem[{{Kong} \& {Ho}(2018)}]{2018ApJ...859..116K}
{Kong}, M., \& {Ho}, L.~C. 2018, \apj, 859, 116

\bibitem[{{Kormendy} \& {Ho}(2013)}]{2013ARA&A..51..511K}
{Kormendy}, J., \& {Ho}, L.~C. 2013, \araa, 51, 511

\bibitem[{{Kroupa}(2001)}]{2001MNRAS.322..231K}
{Kroupa}, P. 2001, \mnras, 322, 231

\bibitem[{{Lamastra} {et~al.}(2009){Lamastra}, {Bianchi}, {Matt}, {Perola},
  {Barcons}, \& {Carrera}}]{2009A&A...504...73L}
{Lamastra}, A., {Bianchi}, S., {Matt}, G., {et~al.} 2009, \aap, 504, 73

\bibitem[{{Madau} \& {Dickinson}(2014)}]{2014ARA&A..52..415M}
{Madau}, P., \& {Dickinson}, M. 2014, \araa, 52, 415

\bibitem[{{Magorrian} {et~al.}(1998){Magorrian}, {Tremaine}, {Richstone},
  {Bender}, {Bower}, {Dressler}, {Faber}, {Gebhardt}, {Green}, {Grillmair},
  {Kormendy}, \& {Lauer}}]{1998AJ....115.2285M}
{Magorrian}, J., {Tremaine}, S., {Richstone}, D., {et~al.} 1998, \aj, 115, 2285

\bibitem[{{Moustakas} {et~al.}(2006){Moustakas}, {Kennicutt}, \&
  {Tremonti}}]{2006ApJ...642..775M}
{Moustakas}, J., {Kennicutt}, Jr., R.~C., \& {Tremonti}, C.~A. 2006, \apj, 642,
  775

\bibitem[{{Nakajima} \& {Ouchi}(2014)}]{2014MNRAS.442..900N}
{Nakajima}, K., \& {Ouchi}, M. 2014, \mnras, 442, 900

\bibitem[{{Nascimento} {et~al.}(2019){Nascimento}, {Storchi-Bergmann},
  {Mallmann}, {Riffel}, {Ilha}, {Riffel}, {Rembold}, {Schimoia}, {da Costa},
  {Maia}, \& {Machado}}]{2019MNRAS.486.5075N}
{Nascimento}, J.~C.~d., {Storchi-Bergmann}, T., {Mallmann}, N.~D., {et~al.}
  2019, \mnras, 486, 5075

\bibitem[{{Noeske} {et~al.}(2007){Noeske}, {Weiner}, {Faber}, {Papovich},
  {Koo}, {Somerville}, {Bundy}, {Conselice}, {Newman}, {Schiminovich}, {Le
  Floc'h}, {Coil}, {Rieke}, {Lotz}, {Primack}, {Barmby}, {Cooper}, {Davis},
  {Ellis}, {Fazio}, {Guhathakurta}, {Huang}, {Kassin}, {Martin}, {Phillips},
  {Rich}, {Small}, {Willmer}, \& {Wilson}}]{2007ApJ...660L..43N}
{Noeske}, K.~G., {Weiner}, B.~J., {Faber}, S.~M., {et~al.} 2007, \apjl, 660,
  L43

\bibitem[{{Obied} {et~al.}(2016){Obied}, {Zakamska}, {Wylezalek}, \&
  {Liu}}]{2016MNRAS.456.2861O}
{Obied}, G., {Zakamska}, N.~L., {Wylezalek}, D., \& {Liu}, G. 2016, \mnras,
  456, 2861

\bibitem[{Oliphant(2006)}]{numpy}
Oliphant, T.~E. 2006, A guide to NumPy, Vol.~1 (Trelgol Publishing USA)

\bibitem[{{Osterbrock} \& {Ferland}(2006)}]{2006agna.book.....O}
{Osterbrock}, D.~E., \& {Ferland}, G.~J. 2006, {Astrophysics of gaseous nebulae
  and active galactic nuclei}

\bibitem[{{Pagel} {et~al.}(1979){Pagel}, {Edmunds}, {Blackwell}, {Chun}, \&
  {Smith}}]{1979MNRAS.189...95P}
{Pagel}, B.~E.~J., {Edmunds}, M.~G., {Blackwell}, D.~E., {Chun}, M.~S., \&
  {Smith}, G. 1979, \mnras, 189, 95

\bibitem[{{Reyes} {et~al.}(2008){Reyes}, {Zakamska}, {Strauss}, {Green},
  {Krolik}, {Shen}, {Richards}, {Anderson}, \&
  {Schneider}}]{2008AJ....136.2373R}
{Reyes}, R., {Zakamska}, N.~L., {Strauss}, M.~A., {et~al.} 2008, \aj, 136, 2373

\bibitem[{{Richstone} {et~al.}(1998){Richstone}, {Ajhar}, {Bender}, {Bower},
  {Dressler}, {Faber}, {Filippenko}, {Gebhardt}, {Green}, {Ho}, {Kormendy},
  {Lauer}, {Magorrian}, \& {Tremaine}}]{1998Natur.395A..14R}
{Richstone}, D., {Ajhar}, E.~A., {Bender}, R., {et~al.} 1998, \nat, 395, A14

\bibitem[{{Salim} {et~al.}(2007){Salim}, {Rich}, {Charlot}, {Brinchmann},
  {Johnson}, {Schiminovich}, {Seibert}, {Mallery}, {Heckman}, {Forster},
  {Friedman}, {Martin}, {Morrissey}, {Neff}, {Small}, {Wyder}, {Bianchi},
  {Donas}, {Lee}, {Madore}, {Milliard}, {Szalay}, {Welsh}, \&
  {Yi}}]{2007ApJS..173..267S}
{Salim}, S., {Rich}, R.~M., {Charlot}, S., {et~al.} 2007, \apjs, 173, 267

\bibitem[{{Salim} {et~al.}(2016){Salim}, {Lee}, {Janowiecki}, {da Cunha},
  {Dickinson}, {Boquien}, {Burgarella}, {Salzer}, \&
  {Charlot}}]{2016ApJS..227....2S}
{Salim}, S., {Lee}, J.~C., {Janowiecki}, S., {et~al.} 2016, \apjs, 227, 2

\bibitem[{{Salpeter}(1955)}]{1955ApJ...121..161S}
{Salpeter}, E.~E. 1955, \apj, 121, 161

\bibitem[{{Scott} \& {Stewart}(2014)}]{2014MNRAS.438.2253S}
{Scott}, A.~E., \& {Stewart}, G.~C. 2014, \mnras, 438, 2253

\bibitem[{{Stern} {et~al.}(2014){Stern}, {Laor}, \&
  {Baskin}}]{2014MNRAS.438..901S}
{Stern}, J., {Laor}, A., \& {Baskin}, A. 2014, \mnras, 438, 901

\bibitem[{{Thomas} {et~al.}(2018){Thomas}, {Kewley}, {Dopita}, {Groves},
  {Hopkins}, \& {Sutherland}}]{2018ApJ...861L...2T}
{Thomas}, A.~D., {Kewley}, L.~J., {Dopita}, M.~A., {et~al.} 2018, \apj, 861, L2

\bibitem[{{Tremonti} {et~al.}(2004){Tremonti}, {Heckman}, {Kauffmann},
  {Brinchmann}, {Charlot}, {White}, {Seibert}, {Peng}, {Schlegel}, {Uomoto},
  {Fukugita}, \& {Brinkmann}}]{2004ApJ...613..898T}
{Tremonti}, C.~A., {Heckman}, T.~M., {Kauffmann}, G., {et~al.} 2004, \apj, 613,
  898

\bibitem[{{Trump} {et~al.}(2015){Trump}, {Sun}, {Zeimann}, {Luck}, {Bridge},
  {Grier}, {Hagen}, {Juneau}, {Montero-Dorta}, {Rosario}, {Brandt},
  {Ciardullo}, \& {Schneider}}]{2015ApJ...811...26T}
{Trump}, J.~R., {Sun}, M., {Zeimann}, G.~R., {et~al.} 2015, \apj, 811, 26

\bibitem[{{Weiner} {et~al.}(2007){Weiner}, {Papovich}, {Bundy}, {Conselice},
  {Cooper}, {Ellis}, {Ivison}, {Noeske}, {Phillips}, \&
  {Yan}}]{2007ApJ...660L..39W}
{Weiner}, B.~J., {Papovich}, C., {Bundy}, K., {et~al.} 2007, \apjl, 660, L39

\bibitem[{{York} {et~al.}(2000){York}, {Adelman}, {Anderson}, {Anderson},
  {Annis}, {Bahcall}, {Bakken}, {Barkhouser}, {Bastian}, {Berman}, {Boroski},
  {Bracker}, {Briegel}, {Briggs}, {Brinkmann}, {Brunner}, {Burles}, {Carey},
  {Carr}, {Castander}, {Chen}, {Colestock}, {Connolly}, {Crocker}, {Csabai},
  {Czarapata}, {Davis}, {Doi}, {Dombeck}, {Eisenstein}, {Ellman}, {Elms},
  {Evans}, {Fan}, {Federwitz}, {Fiscelli}, {Friedman}, {Frieman}, {Fukugita},
  {Gillespie}, {Gunn}, {Gurbani}, {de Haas}, {Haldeman}, {Harris}, {Hayes},
  {Heckman}, {Hennessy}, {Hindsley}, {Holm}, {Holmgren}, {Huang}, {Hull},
  {Husby}, {Ichikawa}, {Ichikawa}, {Ivezi{\'c}}, {Kent}, {Kim}, {Kinney},
  {Klaene}, {Kleinman}, {Kleinman}, {Knapp}, {Korienek}, {Kron}, {Kunszt},
  {Lamb}, {Lee}, {Leger}, {Limmongkol}, {Lindenmeyer}, {Long}, {Loomis},
  {Loveday}, {Lucinio}, {Lupton}, {MacKinnon}, {Mannery}, {Mantsch}, {Margon},
  {McGehee}, {McKay}, {Meiksin}, {Merelli}, {Monet}, {Munn}, {Narayanan},
  {Nash}, {Neilsen}, {Neswold}, {Newberg}, {Nichol}, {Nicinski}, {Nonino},
  {Okada}, {Okamura}, {Ostriker}, {Owen}, {Pauls}, {Peoples}, {Peterson},
  {Petravick}, {Pier}, {Pope}, {Pordes}, {Prosapio}, {Rechenmacher}, {Quinn},
  {Richards}, {Richmond}, {Rivetta}, {Rockosi}, {Ruthmansdorfer}, {Sandford},
  {Schlegel}, {Schneider}, {Sekiguchi}, {Sergey}, {Shimasaku}, {Siegmund},
  {Smee}, {Smith}, {Snedden}, {Stone}, {Stoughton}, {Strauss}, {Stubbs},
  {SubbaRao}, {Szalay}, {Szapudi}, {Szokoly}, {Thakar}, {Tremonti}, {Tucker},
  {Uomoto}, {Vanden Berk}, {Vogeley}, {Waddell}, {Wang}, {Watanabe},
  {Weinberg}, {Yanny}, {Yasuda}, \& {SDSS Collaboration}}]{2000AJ....120.1579Y}
{York}, D.~G., {Adelman}, J., {Anderson}, Jr., J.~E., {et~al.} 2000, \aj, 120,
  1579

\bibitem[{{Zaritsky} {et~al.}(1994){Zaritsky}, {Kennicutt}, \&
  {Huchra}}]{1994ApJ...420...87Z}
{Zaritsky}, D., {Kennicutt}, Jr., R.~C., \& {Huchra}, J.~P. 1994, \apj, 420, 87

\bibitem[{{Zhuang} {et~al.}(2019){Zhuang}, {Ho}, \&
  {Shangguan}}]{2019ApJ...873..103Z}
{Zhuang}, M.-Y., {Ho}, L.~C., \& {Shangguan}, J. 2019, \apj, 873, 103

\end{thebibliography}
\end{document}